 \documentclass[final,5p,times,twocolumn]{elsarticle}

\usepackage{amssymb}
\usepackage{amsfonts}
\usepackage{amsmath}
\usepackage{amssymb}

\journal{Physics Letters B}

\begin{document}

\begin{frontmatter}



\title{A non-abelian quasi-particle model for gluon plasma}


\author{E.P. Politis}
\address{Department of Physics, University of Athens, GR-15874 Athens, Greece}
\author{C.E. Tsagkarakis}
\address{Department of Physics, University of Athens, GR-15874 Athens, Greece}
\author{F.K. Diakonos}
\address{Department of Physics, University of Athens, GR-15874 Athens, Greece}
\author{X.N. Maintas}
\address{Department of Physics, University of Athens, GR-15874 Athens, Greece}
\author{A. Tsapalis}
\address{Hellenic Naval Academy, Hatzikyriakou Avenue, Pireaus 185 39, Greece}
\address{Department of Physics, University of Athens, GR-15874 Athens, Greece}

\begin{abstract}

We propose a quasi-particle model for the thermodynamic description of the gluon plasma which takes into account non-abelian characteristics of the gluonic field. This is accomplished utilizing massive non-linear plane wave solutions of the classical equations of motion with a variable mass parameter, reflecting the scale invariance of the Yang-Mills Lagrangian. For the statistical description of the gluon plasma we interpret these non-linear waves as quasi-particles with a temperature dependent mass distribution. Quasi-Gaussian distributions with a common variance but different temperature dependent mean masses for the longitudinal and transverse modes are employed. We use recent Lattice results to fix the mean transverse and longitudinal masses while the variance is fitted to the equation of state of pure $SU(3)$ on the Lattice. Thus, our model succeeds to obtain both a consistent description of the gluon plasma energy density as well as a correct behaviour of the mass parameters near the critical point.

\end{abstract}


\end{frontmatter}


\section{Introduction}
\label{sec1}
Strongly interacting non-abelian gauge theories are described by $SU(N)$ algebras. These theories possess a strong coupling in the low temperature regime prohibiting perturbative treatment. In addition, the associated degrees of freedom occur exclusively in a confined phase. In higher temperatures, near and above the critical point of 
$SU(N)$ gauge theories, it is expected that the interaction of fermions and gauge fields, namely quarks and gluons in the case of $SU(3)$ color (QCD), is significantly weaker, leading to the deconfined phase known as quark gluon plasma (QGP)\cite{Shuryak1980}. The thermodynamical treatment of QGP has led to the introduction of quasi-particle models (QPMs) \cite{Goloviznin1993, Peshier1994}, primarily aiming to explain the QGP equation of state as obtained from Lattice gauge theory simulations of QCD at finite temperature. In these models and at temperatures higher than the critical value it was assumed that the thermodynamics of a system of interacting massless gluons may be approached by an ideal gas of massive noninteracting gluons. However, the above models \cite{BannurQGP1999} failed to explain the most recent Lattice results \cite{Boyd1996}.

Due to asymptotic freedom, the interaction of quarks and gluons is expected to be very weak at extremely high temperatures. Thus, in these thermodynamic conditions the picture of an ideal gas of (almost) non-interacting particles for the quark-gluon system should provide a good approximation. Nevertheless, as supported by the results of the experiments at the Relativistic Heavy Ion Collider (RHIC) \cite{RHIC}, in the neighbourhood of the critical temperature $T_c$, associated with the transition from hadronic matter to QGP, the interaction is strong and the quark-gluon system is far from the ideal gas scenario \cite{BannurQGP1999,ShuryakQGP2005} sharing features of a perfect fluid \cite{LetessierLQGP2003}. This holds in particular also for the gluon field alone where accurate Lattice results \cite{Boyd1996,Katz2012} demonstrate that the gluon system remains far from the ideal behaviour even for temperatures $5$ times larger than $T_c$.

A useful and common strategy is to restrict the analysis to the gluonic sector considering the emergence of a (non-ideal) gluon plasma above the associated critical temperature. To capture this non-ideal behaviour the QPMs \cite{Peshier,Levai,Kampfer,Schneider,Castorina,Meisinger,Giacosa,Plumari,Filinov,Buisseret,Cao} introduce temperature dependent parameters which are suitably adjusted in order to fit the existing Lattice results. A basic assumption of these models is the presence of a temperature dependent mass for the gluons, a property which may lead to thermodynamic inconsistencies \cite{Yang1995} within the Landau statistical approach \cite{Landau}. This is due to the fact that the temperature dependent mass becomes a thermodynamic quantity affecting the usual relations connecting pressure with energy density. This inconsistency may be healed  by introducing an appropriate constraint in models involving vacuum energy $B(T)$ \cite{Yang1995}. However, a much more natural way to overcome this issue is to use the Pathria \cite{Pathria} approach starting the calculations of thermodynamic quantities and equation of state from the energy density instead of the pressure \cite{Bannur2007TI}.  The pressure is obtained via the integration of a fundamental thermodynamic relation, taking into account the temperature dependence of the gluon mass \cite{Bannur2007TI}. The latter is determined by fitting the Lattice results for the equation of state of pure $SU(3)$ at finite temperature \cite{Boyd1996}. With the  suitable temperature dependence for the gluon mass, the description of Lattice data with such an improved QPM turns out to be quite satisfactory. However, despite of being free from thermodynamic inconsistencies QPMs still include controversial issues from the physical point of view. The use of massive gluons and their treatment as free particles is not fully justified. To resolve this issue, a recent work \cite{Bannur2007} assumed that the gluon mass  emerges through the propagation of gluons in a plasma environment as a collective effect and is furthermore related to the associated plasma frequency. Lattice results for the equation of state are sufficiently reproduced with the use of a single temperature dependent parameter. Regarding the origin of gluon mass, a consistent interpretation is possible in terms of classical solutions of the corrensponding gauge theory equations of motion (e.g see \cite{Smilga} and references therein). In particular, \cite{Savvidy} derives a class of nonlinear plane wave $SU(2)$ solutions obeying a massive relativistic dispersion relation. The mass parameter is free to vary as a consequence of the scale invariance of the Yang-Mills Lagrangian. Despite the fact that such a scenario is based on fundamental properties of Yang-Mills dynamics it has not been associated with the gluon mass in QPMs up to now.

A novel aspect in the framework of QPMs emerged by recent Lattice $SU(3)$ calculations of the temperature dependence of the dynamically generated gluon mass by estimating the inverse gluon propagator in the infra-red limit \cite{Silva2014}. These calculations obtained the gluon mass for a temperature regime just above the critical point, providing further constraints in the phenomenology of QPMs. There are some important consequences of the Lattice results: firstly the dynamical masses of the transverse and longitudinal gluon degrees of freedom differ at low temperatures. Secondly both masses (longitudinal and transverse) behave smoothly as a function of the temperature just beyond the critical point. Thus, the stiff increase of quasiparticle masses as the critical region is approached from above, is incompatible with the Lattice \cite{Silva2014}. A solution to this controversy has been proposed in the framework of QPMs through a Polyakov loop coupling to the quasiparicles \cite{Polyakov}.  

In the present work we develop a QPM which takes into account the most recent Lattice results for the temperature dependence of the dynamical gluon mass \cite{Silva2014} and at the same time is in consistency with the older Lattice calculations of the equation of state \cite{Boyd1996}. The proposed model suggests that the non-linear plane wave solutions of the equations of motion correspond to quasi-particles with variable mass. We develop such a scenario based on a subset of classical solutions for the gauge field, namely those which originate from the $SU(2)$-sector contained in $SU(3)$, assuming that the main characteristics of its non-abelian character are captured by this class. We demonstrate that the non-abelian character of the gluons introduces significant changes in their thermodynamical treatment which are taken into account in the proposed QPM. In contrast to other statistical models which use glueballs with a discrete mass spectrum \cite{Panero} here we assume a continuously varying gluon mass characterized by a specific probability density. To take into account the difference between transverse and longitudinal masses, as calculated from the Lattice, we use two different mass distributions for transverse and longitudinal gluonic degrees of freedom respectively. Furthermore, since the gluon mass distributions are unknown we choose truncated Gaussians (negative mass values are excluded) in each case. The Lattice results (extrapolating to high temperatures when necessary)\cite{Silva2014} are employed to fix the temperature depended mean transverse and longitudinal masses. To reduce the number of free parameters we further assume that the variation is the same in both quasi-Gaussian distributions. Thus, the temperature depended variance is the only free parameter in our model which is determined via a fit to the Lattice results for the $SU(3)$ equation of state \cite{Boyd1996}. The main success of the proposed non-abelian quasi-particle model (NAQPM) is the very good description of two different Lattice results for the gluon plasma using a single free parameter.

The paper is organized as follows. In Section II we present the non-linear plane wave solutions of the $SU(2)$ Yang-Mills theory and we reveal their properties which are relevant for the subsequent formulation of the NAQPM. In section III we introduce the NAQPM for the gluon plasma and the corresponding statistical treatment. In section IV we discuss our results concerning the dependence of the parameters of the proposed model on temperature as well as their compatibility with existing Lattice results. Finally, in section V we give our concluding remarks.

\section{Non linear plane waves in $SU(2)$ Yang-Mills theory}

A simplified description of the gluon field, capturing the basic phenomenological characteristics, can be obtained by the classical $SU(2)$ Yang-Mills theory. Since $SU(2)$ is a subgroup of $SU(3)$, we will restrict our analysis to this case which is easier to handle and at the end of this section we will give some arguments supporting that our treatment can be transferred to the more realistic description with colored $SU(3)$ Yang-Mills.
Neglecting fermionic (matter) degrees of freedom the Lagrangian density of this model is written as:
\begin{equation}
\mathcal{L}=-\frac{1}{4}\mathcal{F}_{\mu \nu}^{a} \mathcal{F}^{\mu \nu a}
\label{eq:eq1}
\end{equation}
where $\mathcal{F}_{\mu \nu}^a$ is the antisymmetric field tensor $\mathcal{F}_{\mu \nu}^{a}=\partial_{\mu}{A_{\nu}^a}-\partial_{\nu}{A_{\mu}^a-g\cdot\epsilon_{a b c}A_{\mu}^bA_{\nu}^c}$ of the gauge field $\mathcal{A}_{\mu}^a$.  The corresponding classical equations of motion for the gauge field are:
\begin{equation}
\partial_{\mu} \mathcal{F}_{\mu \nu}^a +g \cdot \epsilon_{a b c} \cdot A_{\mu}^b \mathcal{F}_{\mu \nu}^c=0
\label{eq:eq2}
\end{equation}
The equations (\ref{eq:eq2}) permit plane wave solutions of the form:
\begin{equation}
A_{\mu}^a=\frac{1}{g} \cdot \varepsilon^a_\mu \cdot \Phi(\omega \cdot t- \vec{k} \cdot \vec{x} )
\label{eq:eq3}
\end{equation}
where $\Phi$ is a scalar function of the plane wave phase 
\begin{equation}
\xi =\omega \cdot t- \vec{k} \cdot \vec{x}
\label{eq:eq4} 
\end{equation}
and $\varepsilon^a_\mu$  (for $a=1,2,3$ and $\mu=0,1,2,3$) is given by
\begin{equation}
\varepsilon^a_\mu=\begin{pmatrix}
-k_1 & -k_2 & -k_3 \\ m + \frac{k_1 ^2}{\omega + m } & \frac{k_1 \cdot k_2}{\omega + m } & \frac{k_1 \cdot k_3}{\omega + m } \\ \frac{k_1 \cdot k_2}{\omega + m } &  m + \frac{k_2 ^2}{\omega + m } & \frac{k_2 \cdot k_3}{\omega + m } \\ \frac{k_1 \cdot k_3}{\omega +  m } & \frac{k_2 \cdot k_3}{\omega + m } & m + \frac{k_3 ^2}{\omega + m }
\end{pmatrix}
\label{eq:eq5}
\end{equation}
The momentum four-vector $\Bbbk^{\mu} = (\omega, \vec{k})$ satisfies the dispersion relation 
$ \omega^2 =\vec{k}^2 +m^2 $ for an arbitrary mass parameter $m$.
Note that the three columns ($a=1,2,3$) of the above matrix~(\ref{eq:eq5}) are 
non-other than the three orthonormal spacelike vectors~\footnote{ Note that $ \varepsilon^a_\mu \varepsilon^{b \mu} = -m ^2 \delta ^{a b} $
 holds.}, orthogonal also to the timelike 
four-momentum vector $\Bbbk_{\mu}$. As such, the solution obeys automatically the 
Lorentz gauge condition $\partial_{\mu} A^{\mu a}=0$ since for plane waves it becomes 
 equivalent to the transversality condition $\Bbbk_{\mu} A ^{\mu a} =0$.

The solution (\ref{eq:eq3}) is most conveniently derived by the authors of \cite{Savvidy} on the proper time frame $\Bbbk^{\mu} = (m, \vec{0})$  with the gauge fixing condition $A_o^a = 0$. On that frame, $A_1^1 = A_2^2 = A_3 ^3 = (m/g) \Phi(\xi)$ with all other components equal to zero and $\Phi(\xi)$ satisfying the equation:
\begin{equation}
\Phi''(\xi)+2 \cdot \Phi(\xi)^3 = 0 \;.
\label{eq:eq6}
\end{equation}
The above equation possesses solutions of the form:
\begin{equation}
\Phi(\xi)= sn[\xi;-1]= cn[\sqrt{2}\xi;\frac{1}{2}]
\label{eq:eq7}
\end{equation}
where $sn[\xi; k^2] (cn[\xi;k^2])$ is the Jacobi elliptic sine (cosine) function with elliptic modulus $k$ \cite{Abram}. 
Thus, the solutions (\ref{eq:eq5}) describe periodic (anharmonic) plane waves with period $\mathcal{P} = 4 K(k)/\sqrt{2}$ where $K(k)$ is the complete elliptic integral of the first kind. For $k^2=1/2$, the period numerically becomes $\mathcal{P}=5.244$. A Lorentz boost with parameters $\gamma=\omega/m , \;\vec{\beta}=\vec{k}/\omega$ leads immediately to the general form (\ref{eq:eq3}). 
The origin of the mass of the gluon field $m$ is traced in the non-linear terms in the Lagrangian (\ref{eq:eq1}) due to the gluon self-interaction. Furthermore, $m$ is free to take any positive value, reflecting the scale invariance of Eq.~(\ref{eq:eq1}). Thus, although scale-free, the Yang-Mills  classical solutions depend on an arbitrary mass-scale $m$ due to the nonlinearity of the theory.  

Solutions of the form of Eq.~(\ref{eq:eq7}) are also found in the $\phi^4$ scalar field theory forming a complete (non-orthogonal) basis \cite{Himpsel2011} for this system. Since there is a mapping of Yang-Mills $SU(N)$ to the scalar $\phi^4$ theory \cite{Frasca2008} one may argue that the solutions in Eq.~(\ref{eq:eq7}) provide also a non-orthogonal basis for $SU(3)$ Yang-Mills gauge theory. The question is if and how the non-linear plane waves (\ref{eq:eq7}) 
can capture the main features of gluon plasma thermodynamics.
Undoubtedly they will influence the counting of gluon microstates. The period of the non-linear plane waves is not $2 \pi$ but $\mathcal{P}=5.244$, thus the number of such stationary states fitting in a fixed volume $V$ is greater that the corresponding number for linear plane waves by a factor $\left(\frac{2 \pi}{\mathcal{P}}\right)^3$. 

\section{The NAQPM}

Let us now proceed with the formulation of the NAQPM as discussed in the introduction. It is useful to list the main assumptions of the model:
\begin{itemize}
\item The microstates of the gluon field at thermal equilibrium consist from non-linear plane wave solutions of the $SU(2)$ classical equations of motion with period $\mathcal{P}=5.244$. Such classical field configurations in quark-gluon plasma studies have been used before \cite{Blaizot1994}. Within our treatment the classical $SU(2)$ non-linear plane waves correspond to massive quasi-particles with a variable mass $m$. 
Since the $SU(3)$ algebra contains three $SU(2)$ subspaces, there exist three 
different ways to embed the above $SU(2)$ solution in $SU(3)$.
The vanisihing trace condition reduces by one the diagonal degrees of freedom, thus we
effectively identify eight degrees of freedom, in consistency with the full $SU(3)$ case. In a more precise sense, the NAQPM does not consider the entire  $SU(3)$ solution space but rather the locally isomorphic case of $SU(2) \times SU(2) \times SU(2)/U(1)$ algebra. Nevertheless, this space contains the basic ingredient of the non-abelian character, namely the non-linearity due to the gauge field self-interaction. A recent publication \cite{Tsapalis2016} has found a larger class of $SU(3)$ plane wave solutions which could in principle be included in the present model. However, these solutions possess an infinite countable set of periods in contrast to the solutions in the $SU(2)$ subspace which have fixed period $\mathcal{P}=5.244$. Thus, their inclusion is highly non-trivial and goes beyond the scope of the present work.

\item Lattice calculations provide evidence that the transverse and longitudinal gluonic degrees of freedom acquire temperature depended masses which differ at low temperatures and approach each other as the temperature increases \cite{Silva2014}. For sufficiently high temperatures these masses attain a common asymptotic linear dependence on the temperature in accordance with perturbation theory predictions \cite{Kapusta}. In the NAQPM this information is incorporated assuming quasi-Gaussian (with restriction to positive values only) distributions for the gluon mass $m$ which have well determined but different, temperature dependent, mean values $\mu_{tr}(T) \sim \langle m \rangle_{tr}$ and $\mu_{lo}(T) \sim \langle m \rangle_{lo}$ for the transverse (tr) and longitudinal (lo) gluonic degrees of freedom respectively. In each case the average is meant over the microstates of the corresponding gluon components (transverse or longitudinal) with temperature depended weights. In such a description the associated variances $\sigma_{tr}\sim \sqrt{\langle m^2 \rangle_{tr} - \mu_{tr}^2}$ and $\sigma_{lo}\sim \sqrt{\langle m^2 \rangle_{lo} - \mu_{lo}^2}$ depend also on temperature and present mass related response functions of the gluonic system. 

\item Within the NAQPM the temperature depended parameters $\mu_{tr}$ and $\mu_{lo}$ are fixed by the recent lattice results \cite{Silva2014} on the gluon propagator. Since we need to calculate thermodynamic properties of the gluon system also for temperatures beyond those investigated in \cite{Silva2014}, we will appropriately extrapolate $\mu_{tr}$ and $\mu_{lo}$. This extrapolation takes into account that in the high temperature regime both quantities approach each other and attain asymptotically a linear temperature dependence as dictated by perturbation theory \cite{Kapusta}. The variances $\sigma_{tr}$ and $\sigma_{lo}$ are the free parameters of the model which are determined in order to fit the Lattice results on the equation of state of the $SU(3)$ Yang-Mills theory \cite{Boyd1996}. To reduce the number of free parameters we further assume that $\sigma_{tr}(T)=\sigma_{lo}(T)\equiv \sigma(T)$. We confirmed numerically that allowing $\sigma_{lo}(T) \neq \sigma_{tr}(T)$ does not influence the results presented in the next section.

\end{itemize}

To calculate the thermodynamic properties of the gluon system within the NAQPM we follow the procedure described in \cite{Bannur2007} to avoid thermodynamic inconsistencies. The energy density is the sum of two contributions, coming from the transverse and longitudinal degrees of freedom respectively, given by ($\beta = 1/T, k_B = 1$):
\begin{equation}
 \epsilon= \epsilon_{tr} + \epsilon_{lo}
 \label{eq:eq8} 
\end{equation}
where
\begin{eqnarray}
\epsilon_i &=& \int_0^{\infty} dm \quad \mathcal{N}_i(\mu_{i},\sigma)~\exp[-\frac{(m-\mu_{i})^2}{2{\sigma}^2}] \nonumber \\
&& \cdot \int d^3 \vec{k} \quad \frac{g_{f,i}}{\mathcal{P}^3} \cdot \frac{\omega(\vec{k},m)}{e^{\beta \omega} -1}~~~~~;~~~~~i=tr,~lo  
\label{eq:eq9}
\end{eqnarray}
with
\begin{multline}
g_{f,tr}=2 \cdot 8~~~~~;~~~g_{f,lo}(k)=8\cdot (1-\frac{{\vec{k}}^2}{{\vec{k}}^2+m^2})
\end{multline}
and
\begin{multline}
\mathcal{N}_i \equiv \mathcal{N}_i(\mu_{i},\sigma)=\frac{1}{\sigma} \sqrt{\frac{2}{\pi}}\cdot \left({1+Erf(\frac{\mu_i}{\sqrt{2}\cdot \sigma})}\right)^{-1}~~~~~;~~~~~i=tr,~lo 
\nonumber \\
\end{multline}

Counting the microstates we have included the prefactor $\left(\displaystyle{\frac{2 \pi}{\mathcal{P}}}\right)^3$ and $\displaystyle{\mathcal{N}_{tr}}$, $\displaystyle{\mathcal{N}_{lo}}$ are normalization factors for the truncated normal distributions describing the gluon mass fluctuations for the transverse and longitudinal degrees of freedom respectively. In the counting of the number of degrees of freedom, we introduce a momentum depending factor $g_{f,lo}(k)=8(1-\displaystyle{\frac{k^2}{k^2 + m^2})}$ for the longitudinal component which takes into account the fact that the longitudinal degrees of freedom vanish for large momenta. For the transverse degrees of freedom we use $g_{f,tr}=2 \cdot 8=16$. In both cases the $8$ corresponds to the eight colored gluons. 

Setting $k=T \cdot x$ \quad transverse and longitudinal energy contributions respectively become:
\begin{multline}
\epsilon_{tr} =\frac{64 \pi \cdot \mathcal{N}_{tr} \cdot T^4}{{\mathcal{P}}^3} \int_0^{\infty} dm \;\; \exp[-\frac{(m-\mu_{tr})^2}{2{\sigma}^2}] \\ 
\cdot \int _0 ^\infty dx \quad x^2 \cdot \frac{\sqrt{x^2 + (m/T)^2}}{e^{\sqrt{x^2+(m/T)^2}}-1}
\nonumber
\end{multline}

\begin{multline}
\epsilon_{lo} =\frac{32 \pi \cdot \mathcal{N}_{lo} \cdot T^4}{{\mathcal{P}}^3} \int_0^{\infty} dm \;\; \exp[-\frac{(m-\mu_{lo})^2}{2{\sigma}^2}] \\ 
\cdot \int _0 ^\infty dx \quad x^2 \cdot (1-\frac{x^2}{x^2 + (m/T)^2}) \cdot \frac{\sqrt{x^2 + (m/T)^2}}{e^{\sqrt{x^2+(m/T)^2}}-1}
\nonumber
\end{multline}

Substituting $x=\frac{m}{T}\cdot sinh(t)$ and using known properties of the Bessel functions, the integration over $x$ is straightforward, leading to the expressions:
\begin{multline}
\epsilon_{tr}= \frac{64 \pi \cdot \mathcal{N}_{tr} \cdot T^4}{{\mathcal{P}}^3} \sum _{l=1} ^ {\infty} \int_0^{\infty} dm \;\; \exp[-\frac{(m-\mu_{tr})^2}{2 {\sigma}^2}]\cdot \frac{1}{l^4} \\
\cdot \left[3(\frac{m}{T}l)^2 \cdot K_2(\frac{m}{T}l) + (\frac{m}{T}l)^3 \cdot K_1(\frac{m}{T}l)\right] \nonumber \\
\end{multline}

\begin{multline}
\epsilon_{lo}= \frac{32 \pi \cdot \mathcal{N}_{lo} \cdot T^4}{{\mathcal{P}}^3} \sum _{l=1} ^ {\infty} \int_0^{\infty} dm \;\; \exp[-\frac{(m-\mu_{lo})^2}{2 {\sigma}^2}]\cdot \frac{1}{l^4} \\
\cdot (\frac{m}{T}l)^3 \cdot K_1(\frac{m}{T}l)
\label{eq:eq10}
\end{multline}
with $K_\nu$ the modified Bessel functions \cite{Abram}.
The mass integral is performed numerically. Having calculated the energy density,
 the pressure is obtained \cite{Bannur2007} integrating the thermodynamically consistent relation  $\displaystyle{\epsilon = T \frac{\partial P}{\partial T}-P}$:
\begin{equation}
\frac{P}{T}=\frac{P_o}{T_o} +\int_{T_o} ^{T} dT ~\frac{\epsilon}{T^2}
\label{eq:eq11}
\end{equation}
The integral in Eq.~(\ref{eq:eq11}) is also performed numerically. The final result for the energy density as well as the pressure, depends of course on the parameters $\mu_{tr}(T)$, $\mu_{lo}(T)$ and $\sigma(T)$.

\section{Numerical results}

Using equations (\ref{eq:eq8},\ref{eq:eq10},\ref{eq:eq11}) we calculate the energy density and pressure for the gluon system at finite temperature. For temperatures $T/T_c \lesssim 1.85$ we use the lattice data presented in \cite{Silva2014} for the transverse and longitudinal mean masses employing linear interpolation to determine their values in between. Beyond $T/T_c \approx 1.85$ longitudinal and transverse masses approach each other and their common behaviour in this high temperature regime becomes linear following the relation:
\begin{equation}
\mu(T)/T_c=1.18 (T/T_c) + 0.48 
\label{eq:eq12}
\end{equation}
\noindent
obtained by a linear fit to the lattice data \cite{Silva2014} in the region $T > 1.5$, with 
$T_c \approx 270~MeV$. The variance $\sigma$ is determined by a Monte-Carlo search such that the lattice data for the energy density and pressure are fitted within $10^{-4}$ accuracy~\footnote{In fact the results are quite robust. Changing the convergence criterion from $10^{-4}$ to $10^{-3}$ leads to a change in the second decimal digit for the parameter $\sigma$.}. Since there are two sets of lattice results for the gluon energy density density and pressure given in \cite{Boyd1996} and \cite{Katz2012} we have fitted both of them. In \cite{Boyd1996} the equations of state are calculated for a relatively small temperature range above $T_c$ while in \cite{Katz2012} the results extend to the region of very high temperatures. Close to the critical region there is a small deviation between the two lattice calculations probably due to finite size effects. Within our approach this deviation is reflected in the values of $\sigma(T)$ necessary to fit the corresponding lattice results.  

In order to guide the search for the $\sigma(T)$ profile so that an approximate continuous function emerges, we use the following strategy:
Assume that we want to determine the parameter $\sigma$ of our model for an (ordered) sequence of temperatures $T_i$, $i=1,2,..,N$ with $T_{i+1} > T_i$. Furthermore, assume that we have located the optimal value $\sigma(T_i)$ for the temperature $T_i$. To estimate the value 
$\sigma(T_{i+1})$ at the subsequent temperature $T_{i+1}$ we explore a region centered at 
$\sigma(T_i)$ and extending up to $50\%$ around it in each direction. This procedure turns out to converge surprisingly fast to the optimal value. This is crucial since the numerical calculations of the integrals in Eqs.~(\ref{eq:eq10}) are performed with high accuracy consuming CPU time. The results of our numerical analysis concerning the equation of state of the gluonic system are presented in Fig.~1.

\begin{figure}[htbp]
\centerline{\includegraphics[width=\columnwidth]{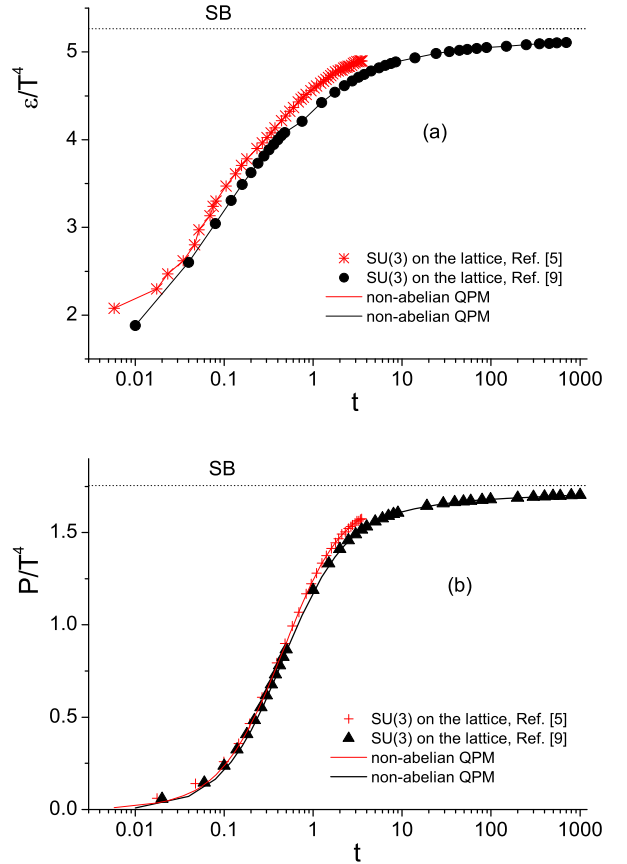}}
\caption{(Color online) (a) The energy density scaled by $T^4$ of the gluon system as obtained by the NAQPM (red and black lines) and the lattice results for the same quantity (red asterisks and black circles). (b) The pressure (also scaled by $T^4$) calculated with the NAQPM (shown with red and black lines) and the corresponding lattice results. The plots display the region $T \geq T_c$ and 
$t$ denotes the reduced temperature $(T-T_c)/T_c$. The lattice results are from \cite{Boyd1996} (red crosses) and \cite{Katz2012} (black triangles). In both plots the dotted line indicates the Stefan-Boltzmann limit.}
\label{fig:fig1}
\end{figure}
 
We observe an excellent agreement between the lattice results given in references
\cite{Boyd1996,Katz2012} and the calculations using the NAQPM introduced in the present work. Note that the agreement is maintained even in the high-temperature regime where the values approach from below the Stefan-Boltzmann limit, indicated by the dotted line in Fig.~1(a,b).
In Fig.~2(a,b) we show the dependence of the parameters $\mu_{tr}$, $\mu_{lo}$ (a) and $\sigma$ (b) on the temperature $T$ as obtained by the Monte-Carlo optimization procedure described above. In Fig.~2a we also display with points the lattice data. We emphasize that the linear dependence of the mass on $T$ for $T/T_c \geq 1.9$, as demonstrated in Fig.~2a, is implied by the lattice calculations in \cite{Silva2014}. Notice that a similar functional form is obtained in high -T perturbative QCD \cite{Kapusta}. The red curve in Fig.~2b is obtained fitting the results of reference \cite{Boyd1996} while the black curve is the NAQPM result for the the description of the corresponding lattice data in reference \cite{Katz2012}. The small shift in $\sigma(T)$ is needed to capture the deviation between the two lattice results. Furthermore, Figs.~2(a) and 2(b) indicate that in the high temperature regime the variance $\sigma$ is proportional to the corresponding common mean mass, as both parameters become linear in T. This is a sign for the presence of a strong correlation between $\mu$ and $\sigma$ and indicates that the mass distribution becomes in fact mono-parametric.  

\begin{figure}[htbp]
\centerline{\includegraphics[width=\columnwidth]{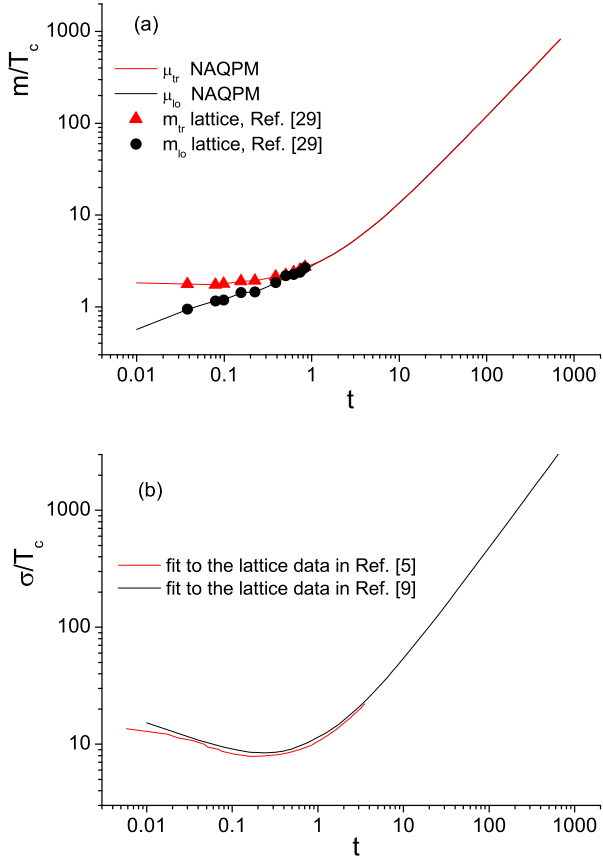}}
\caption{(Color online) (a) The ratios $\frac{\mu_{tr}}{T_c}$ and $\frac{\mu_{lo}}{T_c}$ as a function of the temperature $T$. With points we show the lattice data taken from \cite{Silva2014}. (b) The temperature dependence of the variance $\sigma$ scaled by the critical temperature $T_c$. The plots display the region $T \geq T_c$ and $t$ denotes the reduced temperature $(T-T_c)/T_c$. The black and red lines are obtained through a fit to the results of Ref.~\cite{Katz2012} and Ref.~\cite{Boyd1996} respectively. }
\label{fig:fig2}
\end{figure}

A final comment is in order here. As seen in Fig.~2b, when approaching the critical point an increase of the variance $\sigma$ is observed which does not appear in the plot of the masses shown in Fig.~2a. Thus, within NAQPM criticality is reflected in the temperature dependence of $\sigma$ which defines naturally a measure for the mass fluctuations. This allows for an interpretation of the gluon deconfining transition as a change in the gluon mass spectrum.

\section{Concluding remarks}

We have introduced a quasi-particle model for the thermodynamical description of the gluon plasma which takes non-abelian characteristics into account. In particular, the microstates building up the statistical ensemble for the gluon are determined from non-linear plane wave solutions of the associated classical dynamics, containing a free mass parameter. They correspond to quasi-particles with a continuously varying mass having a quasi-Gaussian distribution with temperature dependent mean mass and variance. To bridge the gap with the Lattice \cite{Silva2014} we use different mean masses for transverse and longitudinal gluonic degrees of freedom keeping a common variance. Tuning appropriately the variance $\sigma$, which is the only free parameter in our model, we reproduce with high accuracy the results of the most recent Lattice calculations for the gluon plasma equation of state. At the same time we avoid singular behaviour of the mean mass close to the critical point, a feature which is common in QPMs but is in contradiction to Lattice results. In our approach the traces of the transition to the gluon plasma phase are imprinted on the rapid increase of the gluon mass variance in the neighbourhood of the critical point. It would be interesting to look for a similar behaviour in the phase diagram when matter degrees of freedom are included. However, this is left for future investigations.





\end{document}